\documentclass[twocolumn,showpacs,preprintnumbers,amsmath,amssymb]{revtex4}

\usepackage{graphicx}
\usepackage{dcolumn}
\usepackage{bm}

\begin{document}


\title{Magnetically-controlled velocity selection in a cold atom
sample using stimulated Raman transitions}

\author{Matthew L. Terraciano}
\author{Spencer E. Olson}%
\author{Mark Bashkansky}%
\author{Zachary Dutton}%
\author{Fredrik K. Fatemi}%

\affiliation{Naval Research Laboratory, 4555 Overlook Ave. S.W.,
Washington, DC 20375}

\date{\today}

\begin{abstract}
We observe velocity-selective two-photon resonances in a cold atom
cloud in the presence of a magnetic field.  We use these resonances
to demonstrate a simple magnetometer with sub-mG resolution. The
technique is particularly useful for zeroing the magnetic field and
does not require any additional laser frequencies than are already
used for standard magneto-optical traps.  We verify the effects
using Faraday rotation spectroscopy.
\end{abstract}

\maketitle

\section{\label{sec:level1}Introduction}

Stimulated Raman transitions that couple atomic ground states with
counterpropagating laser beams are resonant only within a narrow
velocity band. This atomic velocity selection~\cite{Kasevich} has
proven to be a useful tool for a variety of experiments, including
subrecoil Raman cooling~\cite{boyer:043405}, atom
interferometry~\cite{McQuirk}, and atom velocimetry~\cite{Chabe}.
Stray magnetic fields can adversely affect this process by shifting
the magnetic sublevels, thereby perturbing the participating
velocity bands~\cite{boyer:043405, Chabe, Garreau}. Conversely, when
transitions occur between different magnetic sublevels of a single
hyperfine level, velocity selectivity can provide an excellent
measure of stray or applied magnetic fields.

Elimination of stray fields to sub-milliGauss levels is particularly
important for sub-recoil cooling processes~\cite{boyer:043405,
Vuletic}. Typically, these fields are nulled by Helmholtz coils
along each cartesian direction. Correct compensation currents can
roughly be estimated by visual indicators such as atom expansion in
an optical molasses, but these cues are strongly dependent on
optical alignment. Stray fields can be directly measured using, for
example, Faraday spectroscopy, which provides picoTesla
sensitivity~\cite{Isayama, Jessen, Kaiser}, but requires additional
laser frequencies and time-resolved polarimetry. Measurement of
vector magnetic fields with magneto-resistive probes has been used
for active compensation of both DC and AC fields, but needs several
sensors placed externally to the vacuum chamber~\cite{Garreau}.

In this paper, we describe a simple imaging technique for measuring
magnetic fields with sub-milliGauss resolution using a sample of
cold atoms from a point trap. The technique relies on
velocity-selective two-photon resonances \cite{Kasevich} (VSTPR) in
a magnetic field, where the two-photon resonance occurs between
different magnetic sublevels within a single hyperfine level. When
applied to atoms cooled in an alkali-vapor MOT, no additional laser
frequencies are required, because the VSTPR pulse can be derived
from the repumping laser beams.  The two main requirements are 1)
VSTPR beams along a horizontal axis and 2) a CCD camera whose
optical axis is orthogonal to the propagation direction of the VSTPR
beams.

\section{\label{sec:level2}Background}

\begin{figure}
\includegraphics[width=\columnwidth]{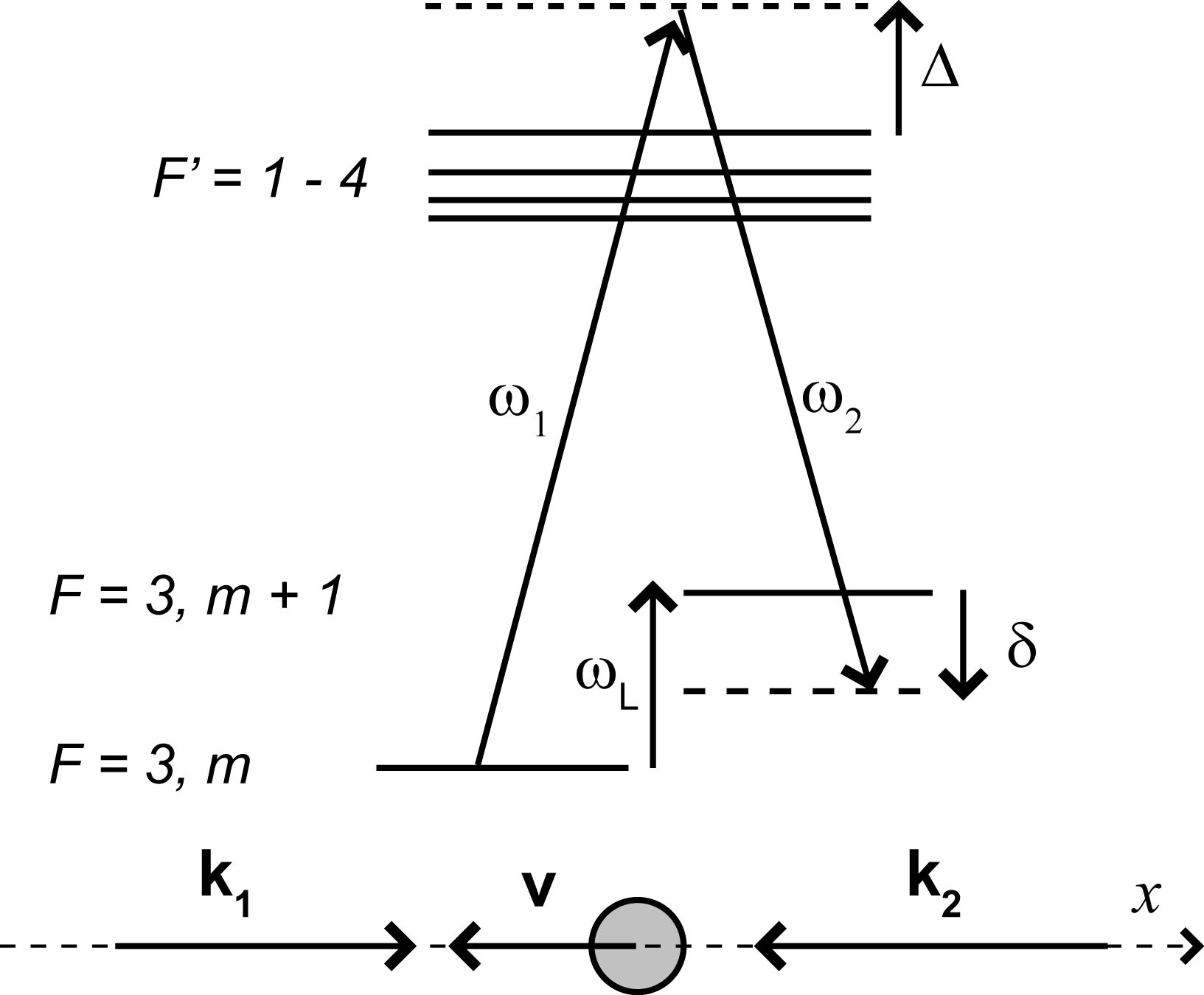}
\caption{\label{fig:EnergyLevelDiagram} Relevant energy levels for
the Raman transitions.  Magnetic sublevels are split by the Larmor
precession frequency, $\omega_L$.  The polarization configuration is
lin~$\perp$~lin.  The energy scale is exaggerated for clarity.}
\end{figure}

In this section, we briefly describe the process.
Figure~\ref{fig:EnergyLevelDiagram} shows the relevant energy
levels. An atom moving with velocity \textbf{v} along the $x$-axis
is exposed to a light field composed of two counterpropagating laser
beams with wave vectors $\mathbf{k_1}$ and $\mathbf{k_2}$, where
$\mathbf{k_1}\simeq-\mathbf{k_2}\simeq\mathbf{k}$ along the
$x$-axis. The polarizations of the two beams are lin~$\perp$~lin.
For an arbitrary B-field, this polarization configuration couples
$m_f - m_i = \Delta{m} = 0, {\pm}1, {\pm}2$ magnetic sublevels of a
single hyperfine level, where we choose our quantization axis along
the magnetic field. Absorption of a photon from one beam and
emission into the other results in a linear momentum change of
$\pm\hbar(\mathbf{k_1} - \mathbf{k_2}) \approx \pm2\hbar\mathbf{k} =
\pm2Mv_r\mathbf{\hat{x}}$, where $v_r$ is the recoil velocity and
$M$ is the mass. The one-photon detuning, $\Delta$ is chosen to be
much larger than the hyperfine splittings of the upper state.

The two-photon detuning is defined here as $\delta = \omega_1 -
\omega_2 - (\Delta{m})\omega_L$, where $\omega_L$ is the Zeeman
splitting. In a small magnetic field, $\hbar\omega_L=g_F{\mu}_B{B}$,
where $g_F$ is the gyromagnetic ratio, and $\mu_B$ is the Bohr
magneton. For $^{85}$Rb, $g_F{\mu}_B/\hbar$ = 466.74
kHz/Gauss~\cite{gF}. Two-photon resonance occurs for atoms whose
velocity satisfies
\begin{equation}
 \delta = \delta_{LS} + \delta_D + 4\delta_{r},
\end{equation}

\noindent where the two-photon Doppler shift, $\delta_D =
2\mathbf{k}{\cdot}\mathbf{v}$, and the recoil frequency $\delta_{r}
= {\hbar}k^2/2M$.  The relative light shift, $\delta_{LS}$ is a weak
function of the participating magnetic sublevels, which are both in
the same hyperfine ground state. Apart from $\delta_{LS}$, and with
$\omega_1 = \omega_2$, the resonance condition is satisfied for
atoms with $\mathbf{v} = \mathbf{v_0}\pm{v_r\mathbf{\hat{x}}}$ such
that $2\mathbf{k}{\cdot}\mathbf{v_0} = (\Delta{m})\omega_L$.

Throughout the duration of the VSTPR pulse, resonant atoms oscillate
between the two momentum states separated by $2{\hbar}\mathbf{k}$.
Because the atoms are initially confined in a point trap, an image
of the atom cloud after expansion is a spatial map of the average
velocity distribution, which has been perturbed by the VSTPR pulse.
A freely expanding cloud has approximately a smooth Gaussian
velocity spectrum along $x$ (the VSTPR beam direction), but the
momentum oscillations that occur for the resonant atoms alter this
average velocity distribution. Images taken along a camera direction
orthogonal to $x$ record these narrow perturbations.

\section{\label{sec:level3}Experiment Setup}

\begin{figure}
\includegraphics{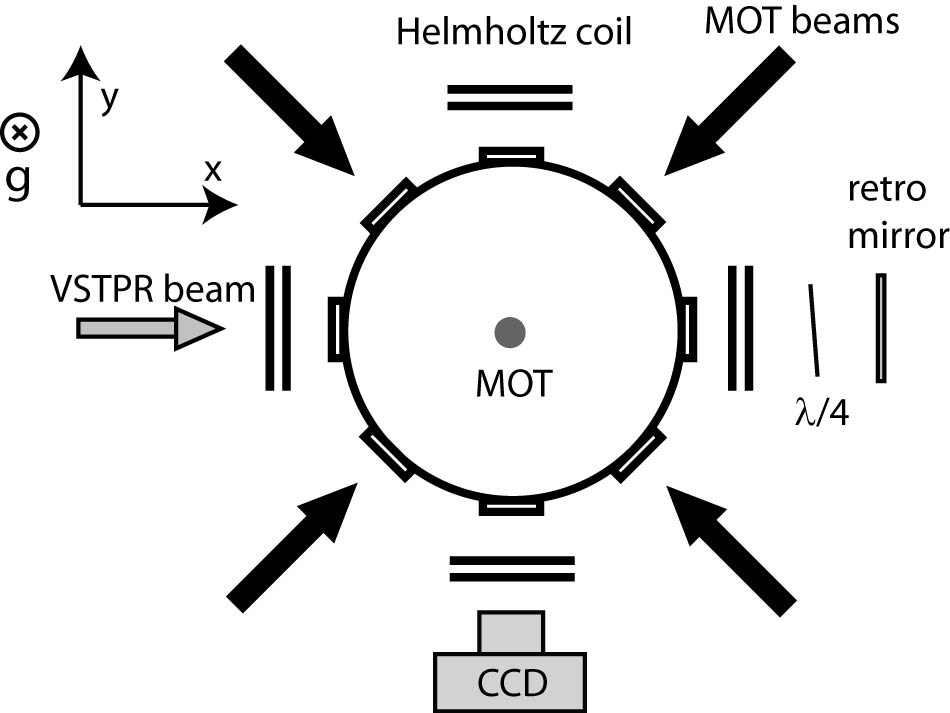}
\caption{\label{fig:chamber} Experimental setup for observing VSTPR
in a magnetic field.}
\end{figure}

The layout of our apparatus is shown in Fig.~\ref{fig:chamber}.  The
experiment begins with a vapor cell MOT containing 10$^7$ $^{85}$Rb
atoms.  The MOT diameter is $\approx$500~$\mu$m and the temperature
is $\approx$200~$\mu$K.  The cooling beams (detuned -18 MHz from the
$F$=3 - $F'$=4 transition) are derived from an extended cavity diode
laser (New Focus Vortex, model 6013) that seeds a 120 mW laser diode
(Sharp GH0781JA2C). 60 mW is directed to one port of a 2 x 3
polarization maintaining (PM) fiber splitter. Each of the three
output fibers carries 12 mW. The outputs are collimated using 100-mm
focal length, 50mm-diameter achromats, giving 1/e$^2$ beam diameters
of 24 mm. The three beams propagate along orthogonal directions and
are retroreflected; one pair is vertical and the other two are in a
horizontal ($x$-$y$) plane. The repump light, connecting $F=2
\rightarrow F'=3$, is derived from an independent Vortex laser that
also seeds a diode. In normal MOT operation, 15 mW of repump light
is coupled into the other port of the 2x3 coupler.

Our VSTPR beam is spatially filtered by PM fiber, and is collimated
by a 60mm gradient-index singlet lens ($1/e^2$ beam waist $\omega_0$
= 7.5 mm).  We use up to 20 mW laser power. It is retroreflected in
a lin $\perp$ lin configuration.  For B-field control, we use three
orthogonal pairs of Helmholtz coils.  The magnetic field at the atom
cloud has components $B_i = \alpha_i(I_i - I_{0i})$ where $\alpha_i$
are the slopes $dB_i/dI_i$, $I_i$ are the applied currents, and
$I_{0i}$ are the currents required for compensation along each
Cartesian direction. The VSTPR beam travels horizontally along the
axis of the $x$-directed coil pair.  For all results in this paper,
we have used a second beam derived from the repump laser as our
VSTPR beam ($\Delta=3$GHz).

At time T = 0, the atoms are released from the MOT by extinguishing
all laser beams and the MOT coils.  The bias magnetic coils remain
on.  We do not perform any molasses cooling, because the large
velocity spread of the hotter sample of atoms provides greater range
over which velocity selection can occur.  At time $T_r \approx 15$
ms, the VSTPR pulse is switched on for 5 ms and at $T_i = 40$ ms,
the MOT cooling and repump beams are switched on to image the
expanded cloud onto the CCD camera.

\section{\label{sec:level4}Results}

\begin{figure}
\includegraphics[width=8.4cm]{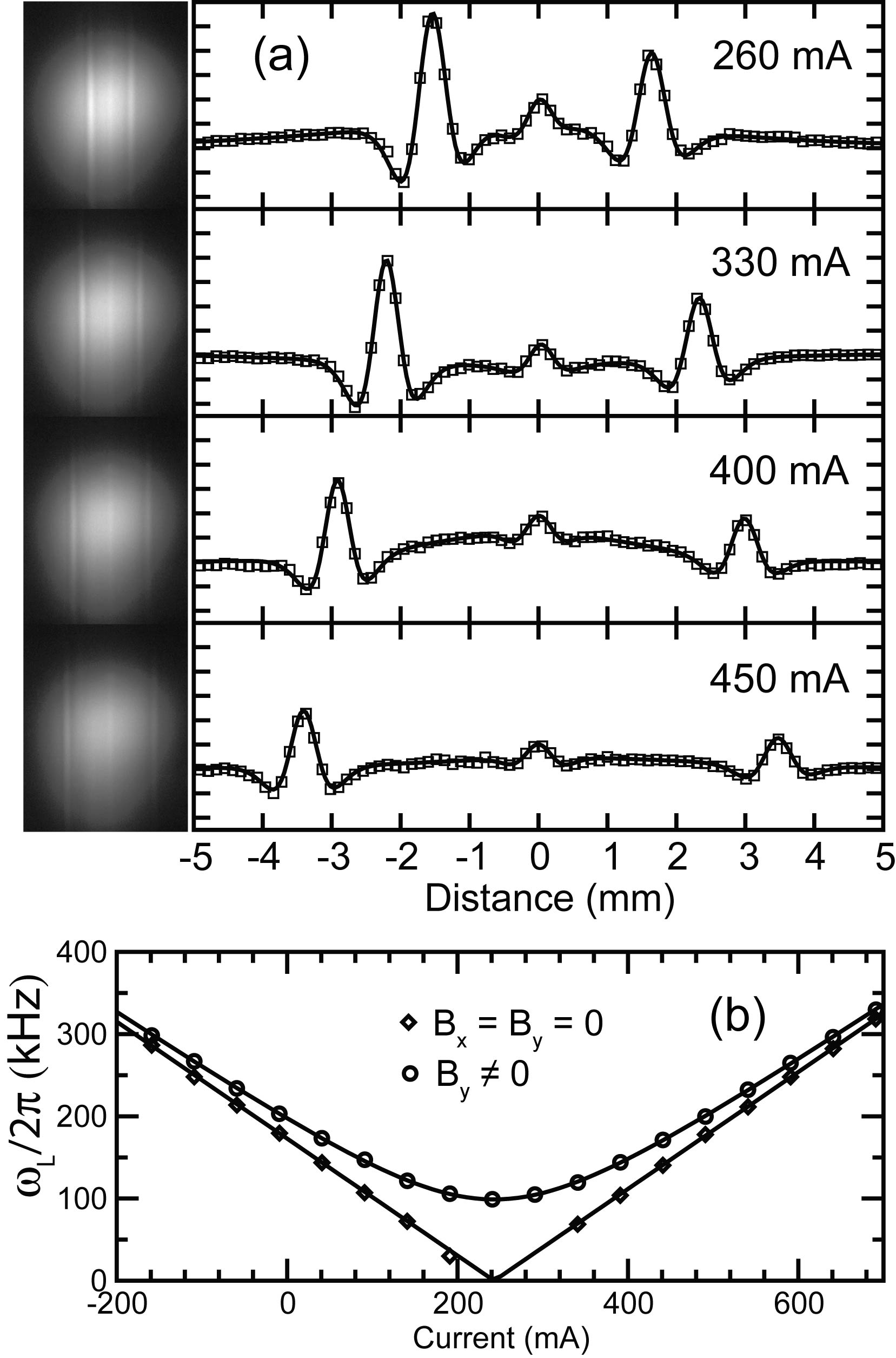}
\caption{\label{fig:stripe_image} a) Left:  Raw images of the
expanded atom cloud at different Helmholtz currents after exposure
to the VSTPR beams. Right: Corresponding cross-sectional profiles
after background subtraction. Fits are shown as solid lines.  b)
Larmor precession frequency extracted from the stripe separation as
a function of current in the $z$-directed coils for a nonzero $B_y$
field (circles) and for $B_x=B_y=0$ (diamonds).  Fits shown as solid
lines.}
\end{figure}

We control the magnetic field by changing the current in the
Helmholtz bias coils.  We first demonstrate the effect by changing
the current in the $z$-directed Helmholtz coils.
 Fig.~\ref{fig:stripe_image}a shows typical images recorded by the
CCD camera for four different current settings.  The VSTPR pulse
creates perturbations that appear as vertical stripes in the
expanded cloud. Also shown in Fig.~\ref{fig:stripe_image}a are the
cross sections after subtracting the images obtained with no VSTPR
pulse. The spatial location of each peak corresponds to the average
velocity class satisfying $2{\mathbf{k}\cdot}\mathbf{v_0} =
\pm(\Delta{m})\omega_L$ that participated in the VSTPR process. In
general, $\omega_L$ is a function of position due to spatially
varying magnetic fields.  These gradients could tip, bend, or blur
the resonant stripe features.  Under normal operating conditions,
however, we do not observe these effects.  With the B-field
primarily along the $z$-axis and using lin~$\perp$~lin
polarizations, the VSTPR pulse mainly connects $\Delta{m} = 1$
transitions. The cross-sectional profile of each vertical stripe can
be estimated by several functional forms. For simplicity, we fit
each peak to two Gaussian profiles such that the total area is 0,
but we describe a more exact fitting function below.

In addition to the peaks corresponding to velocity classes at
$\pm{v_0}$, there is a peak at v = 0 due to $\Delta{m}=0$
transitions that arises from the longitudinal magnetic field
component. This is a useful marker for balancing the overall
velocity distribution. We rarely observe features corresponding to
${\Delta}m = {\pm}2$ transitions. Detailed calculations of the
relative transition strengths will be published elsewhere, but in
general, the strengths of the ${\Delta}m = {\pm}2$ transitions are
roughly two orders of magnitude smaller than those for ${\Delta}m =
0, {\pm}1$ for $\Delta = 3$~GHz.  It is important to note that the
appearance of narrow features in the expanded cloud is not the
result of cooling, because there is no dissipative force. For
smaller $\Delta$, on the order of the hyperfine splittings of the D2
manifold, spontaneous scattering events can lead to
magnetically-induced laser cooling~\cite{Metcalf}.

The $z$-directed Helmholtz bias coils are each made of 24 gauge wire
wound on an 8" vacuum flange, 1" long, and centered 5.8 cm from the
MOT, producing a field of approximately 1.5 G/A.  Our fits to
Fig.~\ref{fig:stripe_image} show that after 35 ms of falling time,
the separations of the stripes for the 4 current settings 260 mA,
330 mA, 400 mA, and 450 mA are 3182(4) $\mu$m, 4538(4), $\mu$m,
5902(4) $\mu$m, and 6873(4) $\mu$m. The listed errors are
statistical. Our leading systematic source of multiplicative error
is the pixel calibration on the camera of $\approx$0.2\%. In
section~\ref{sec:level6}, we briefly describe a second calibration
procedure that removes pixel calibration errors. The 4 $\mu$m error
in stripe separation corresponds to an error of ~300 $\mu$G. Note
that the error in magnetic field will be explicitly dependent on the
atom species through $g_F$ and the VSTPR wavelength.

The scalar magnetic field measurements for several current settings
are shown in Fig.~\ref{fig:stripe_image}b.  Because the stripe
separation is proportional to ${(B_x^2 + B_y^2 + B_z^2)^{1/2}}$, a
plot of $\omega_L$ versus the current in the $z$-directed bias coil
traces a hyperbola, the minimum of which determines the field
component perpendicular to $z$ and the compensation current
$I_{0z}$. We show two cases, one with nonzero $B_y$, and one for the
case in which $B_x, B_y$ have been zeroed. From a fit to these
plots, we extract $\alpha_z = 1.524(2)$ G/A and $I_{0z}$ = 243.1(2)
mA, which corresponds to a compensation level of 300~$\mu$G. In
practice, it is simple to zero the magnetic field by viewing
real-time images of the expanded cloud and adjusting the currents
along each axis for minimum stripe separation.  We note that when
the total magnetic field is close to zero, the stripes begin to
overlap and are no longer resolved. Compensation is achieved when
the overlap is maximized, resulting in a single narrow feature.  In
our experience, this real-time adjustment of the stripe separation
results in compensation to milliGauss levels without any data
analysis.

The visibility of the stripe features is dependent on a few factors.
First, since the image on the CCD camera is a convolution of the
initial MOT size with the velocity distribution, the contrast
increases for trapped samples with smaller physical dimensions.
Optimally, the imaging should be performed after the cloud has
expanded enough that two velocity classes separated by
$2{\hbar}\mathbf{k}/M$ can be resolved. If the initial MOT has a
radius R, this means that the imaging should be performed after a
time $R/v_{rec}$ from the release of the atoms from the trap. In
practice, the features are easily observed with imaging times
significantly less because the effect does not require that the
recoil velocities be resolved, only that perturbations to the
average velocity distribution can be observed.

\begin{figure}
\includegraphics[width=8.4cm]{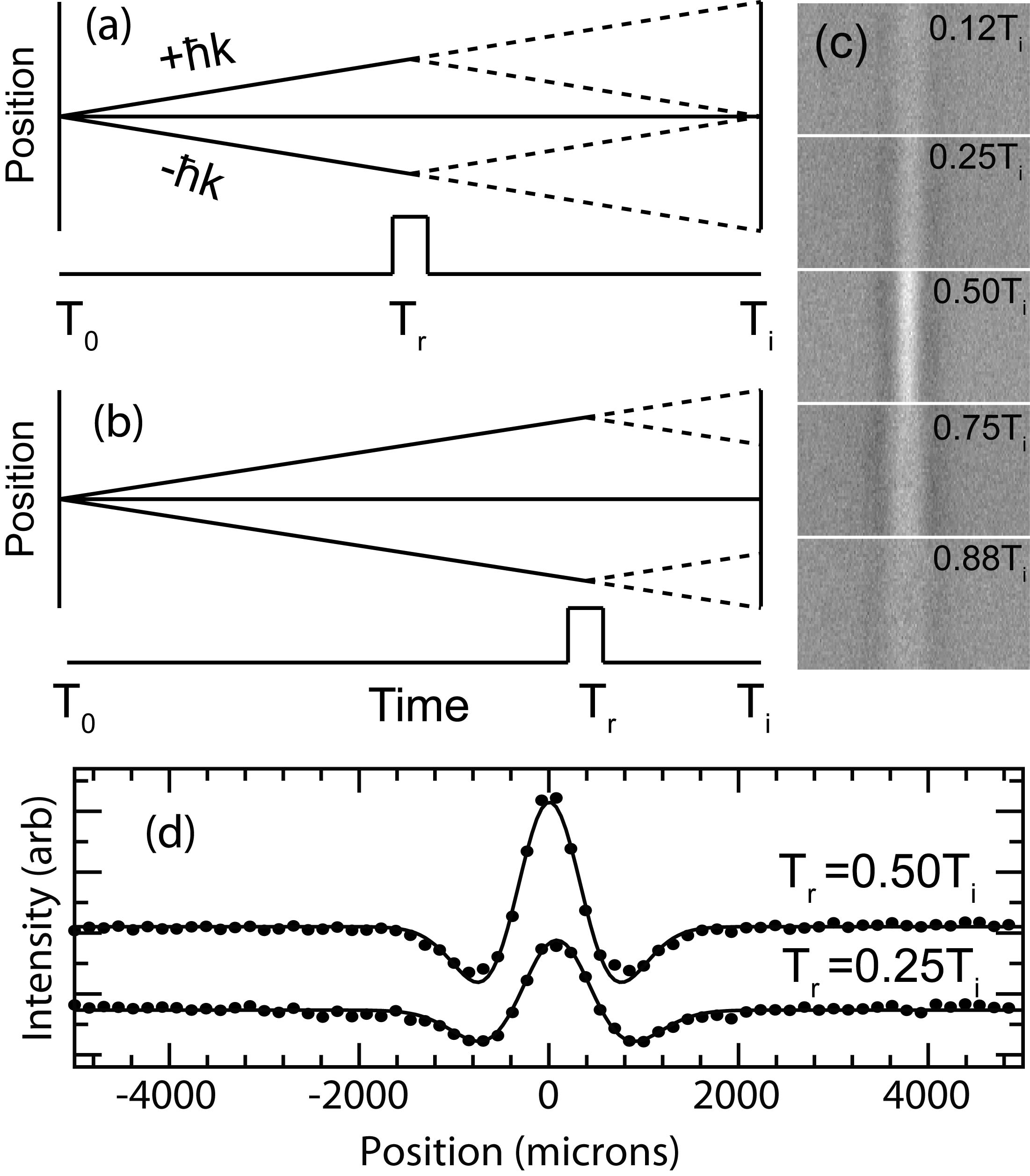}
\caption{ a) Importance of timing of the VSTPR pulse.  An atom
moving with ${\hbar}k$ has its momentum reversed at $T_r$.  If this
pulse occurs at $T_i/2$, atoms initially having ${\pm}{\hbar}k$ are
overlapped at $T_i$. b) If the pulse arrives later or earlier,
contrast is reduced. c) Pictures of a single stripe using different
$T_r$, showing optimum contrast at $T_i/2$. d) Stripe cross-sections
for $T_r=0.5T_i$ and $T_r=0.25T_i$, along with fits using
Equation~\ref{eqn:Gaussian} (solid line).}
\label{fig:TimingSequence}
\end{figure}

Other parameters that control the visibility of the stripes are the
duration and timing of the VSTPR pulse.  Because this measurement is
time-averaged over the duration of the pulse, shorter pulses reduce
blurring effects due to time-varying fields.  Furthermore, if
applied at $T_r = T_i/2$ they can maximize stripe contrast.
Figure~\ref{fig:TimingSequence}a shows the effect on stripe contrast
as a function of $T_r$ for atoms in zero field. An atom initially at
the origin will return to the origin if its momentum is reversed by
a $\pi$-pulse at $T_i/2$.  For $\Delta{T}=T_r-T_i/2\neq0$, the atoms
are still deflected, but the atoms with initial momenta of
${\pm}{\hbar}k$ no longer spatially overlap at $T_i$
(Fig.~\ref{fig:TimingSequence}b). In Fig.~\ref{fig:TimingSequence}c,
we show this effect experimentally using VSTPR pulse durations of
200 ${\mu}$sec at different $T_r$. This simple geometrical picture
suggests an approximate functional form for the
background-subtracted stripe cross section under these conditions:

\begin{eqnarray}
Y(x) & = & G(x_0-2v_r\Delta{T}) + G(x_0+2v_r\Delta{T}) \nonumber\\
     & - & G(x_0-v_rT_i) - G(x_0+v_rT_i)
     \label{eqn:Gaussian}
\end{eqnarray}

\noindent where $G(x_0)$ is a Gaussian centered at $x_0$. In
Fig.~\ref{fig:TimingSequence}d, we show fits when $T_r=0.5T_i$ and
$T_r=0.25T_i$.  For all other data presented in this manuscript, we
used pulse durations of 5 msec.  Although AC magnetic fields were
not compensated, these longer pulses showed no measurable broadening
in our experiment.  The $\pi$-pulse duration depends on the
particular magnetic sublevels involved, so for our spin-unpolarized
sample we generally choose a pulse duration that provides
consistently strong signals over a broad range of Raman pulse
intensities.  For a typical VSTPR beam intensity of 10~mW/cm$^2$,
the resonant 2-photon Rabi frequency is ${\approx}2\pi\times$10~kHz.

\section{\label{sec:level5}Comparison with Faraday Spectroscopy}

\begin{figure}
\includegraphics[width=8.4cm]{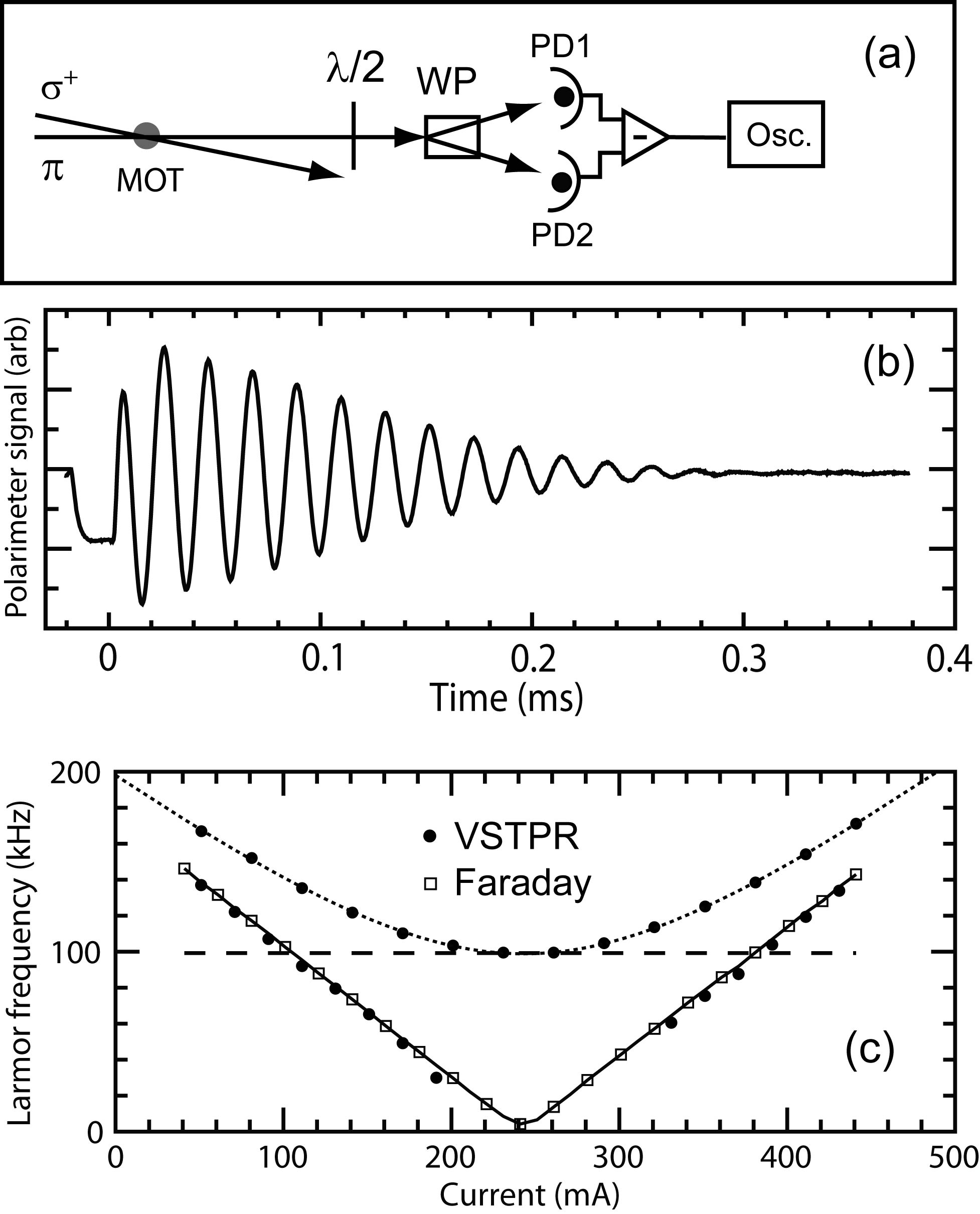}
\caption{\label{fig:faraday} a) Schematic of Faraday spectroscopy
setup. WP: Wollaston prism.  b) Typical Faraday signal. c)
Comparison of magnetic field measurement with VSTPR (circles) and
with Faraday spectroscopy (squares) for similar conditions as
Fig.~\ref{fig:stripe_image}b. Solid line is a fit to the Faraday
data. Dotted line is a fit to the VSTPR data. Dashed line indicates
the minimum of the hyperbola as measured by Faraday spectroscopy.}
\end{figure}

We have verified this VSTPR technique by using Faraday rotation
spectroscopy to measure the magnetic field~\cite{Isayama, Jessen,
Kaiser}. To perform these measurements, an additional pair of laser
beams is used along the $x$-axis (Fig.~\ref{fig:faraday}a).  The
atoms are optically pumped into the $F=3, m_F = 3$ stretched state
(in the $x$-basis) by a 100 $\mu$s $\sigma^+$ pulse connecting $F=3
\rightarrow F' = 3$. This beam contains a small amount of repumper
light to keep the atoms in F=3. When this light is extinguished, the
atoms begin precessing freely.  A linearly polarized probe beam with
$\simeq$100 $\mu$W and waist $\omega_0 = 500{\mu}$m passes through
the atom cloud to a simple polarimeter consisting of a Wollaston
prism that splits the probe beam into two orthogonal polarization
states that are detected by a balanced photodetector \cite{Hobbs}.
We make these measurements at the same time delay as the VSTPR
measurements. A typical Faraday signal is shown in Fig.
\ref{fig:faraday}b.  The comparison of the VSTPR and Faraday
techniques is shown in Fig.~\ref{fig:faraday}c.  Because we cannot
compare Faraday and stripe measurements near zero field where the
stripes are unresolved, we show in Fig.~\ref{fig:faraday}c the
stripe data with and without a transverse field along $y$. This
transverse field allows a stripe measurement at $I_{0z}$, which is
in good agreement with the Faraday measurement (dashed line in
Fig.~\ref{fig:faraday}c). From the Faraday measurements with no
transverse field, also shown in this figure, we derive $\alpha_z =
1.547(3)$ G/A and $I_{0z} = 242.5(1)$ mA, both of which agree well
with the VSTPR technique.

\section{\label{sec:level6}Calibration}

To obtain correct values of the magnetic field, the stripe
separation must be carefully measured.  Without calibration, the
technique is still useful for determining the $I_{0i}$ of each coil
by simply finding the minimum stripe separation, which is
independent of this source of systematic error.  Because this is an
imaging technique, good estimates of spatial calibration can be made
simply by measuring the magnification on the CCD camera if
high-quality imaging lenses are used.  Most lenses exhibit some
degree of aberrations that make accurate pixel calibration difficult
below the 0.5\% level.  Even without this error, other slight
systematic errors, such as the exact functional form used to fit the
stripe cross section may remain.  In this section, we describe a
technique for calibrating the Zeeman shifts indicated by the stripes
in a more direct manner that eliminates these systematic errors.

\begin{figure}
\includegraphics[width=8.4cm]{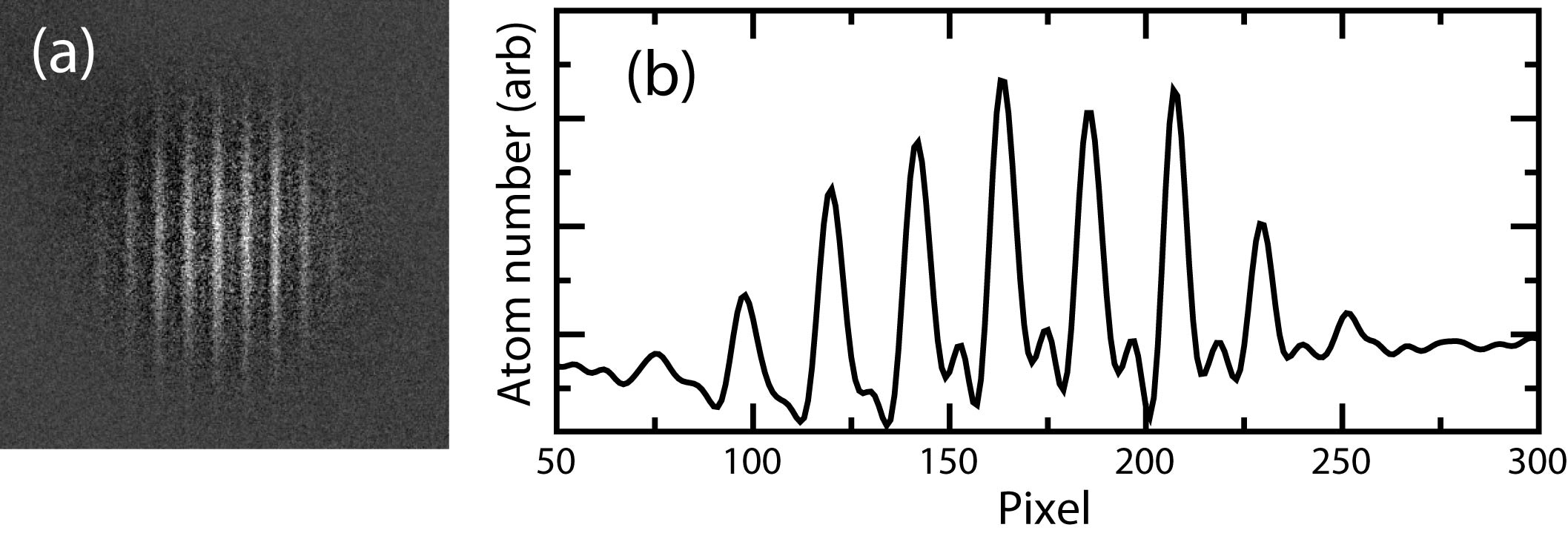}
\caption{\label{fig:calibration}  a) Image of atom cloud (after
background subtraction) taken with several sidebands separated by
100 kHz imposed on one VSTPR beam. b) Cross sectional profile.}
\end{figure}

Instead of relying on an accurate pixel calibration, the splittings
of the stripes can be determined by making $\omega_1 = \omega_2 +
\delta_{12}$, where $\delta_{12}$ is a frequency shift imposed by an
RF source.  The resonant velocity classes are now determined by
$2\mathbf{k}{\cdot}\mathbf{v} = ({\Delta}m)\omega_L + \delta_{12}$.
To demonstrate this idea, the retroreflecting mirror in
Fig.~\ref{fig:chamber} is replaced by a counterpropagating beam of
the same diameter and power.  This counterpropagating beam is
derived from the original, so it is phase locked to $\omega_1$, but
its frequency is shifted by two acousto-optic modulators (AOM) to
achieve small $\delta_{12}~(<$1~MHz$)$.  Additionally, we frequency
modulate one AOM so that its instantaneous frequency is
$\omega_{RF}$ + Asin($\omega_{m}t$), where $\omega_{RF}$ is the
drive frequency, $\omega_{m}$ is the modulation frequency, and A is
the maximum frequency deviation. This imparts frequency sidebands at
$n\omega_{m}$, where $n$ is an integer, so that multiple
$\delta_{12}$ are produced simultaneously. In
Fig.~\ref{fig:calibration}, we show an image and cross section taken
with $\omega_{m} = 2{\pi}\times100$~kHz.  For $\delta_{12}=0$, the
range of measurable Zeeman shifts is limited to the Doppler width of
the atom cloud ($\simeq$1G).  A nonzero $\delta_{12}$ overcomes this
limitation by shifting the stripe to an accessible velocity class.

\section{\label{sec:level7}Conclusion}

We have used velocity-selective resonances between magnetic
sublevels of a single hyperfine level in $^{85}$Rb to measure
magnetic fields in a cold atom cloud.  The resonances are easily
observed with no additional laser frequencies than are required for
MOTs, and can be used to measure magnetic fields with sub-mG
resolution. Because of its simplicity, this technique should prove
especially useful for aiding magnetic field compensation, for which
purpose no calibration is required.

This work was funded by the Defense Advanced Research Projects
Agency and the Office of Naval Research.


\end{document}